\documentclass[]{pasj01}
\draft

\begin{document} 
\Received{}
\Accepted{}

\title{
Wandering of the central black hole in a galactic nucleus and correlation of the black hole mass with the bulge mass
}

\author{Hajime \textsc{Inoue}\altaffilmark{1}}%
\altaffiltext{1}{Institute of Space and Astronautical Science, Japan Aerospace Exploration Agency, 3-1-1 Yoshinodai, Chuo-ku, Sagamihara, Kanagawa 252-5210, Japan}
\email{inoue-ha@msc.biglobe.ne.jp}


\KeyWords{accretion, accretion disk - galaxies: active - galaxies: nuclei} 

\maketitle

\begin{abstract}
We investigate a mechanism for a super-massive black hole at the center of a galaxy to wander in the nucleus region.
A situation is supposed in which the central black hole tends to move by the gravitational attractions from the nearby molecular clouds in a nuclear bulge but is braked via the dynamical frictions by the ambient stars there.
We estimate the approximate kinetic energy of the black hole in an equilibrium between the energy gain rate through the gravitational attractions and the energy loss rate through the dynamical frictions, in a nuclear bulge composed of a nuclear stellar disk and a nuclear stellar cluster as observed from our Galaxy.
The wandering distance of the black hole in the gravitational potential of the nuclear bulge is evaluated to get as large as several 10 pc, when the black hole mass is relatively small.
The distance, however, shrinks as the black hole mass increases and 
the equilibrium solution between the energy gain and loss disappears when the black hole mass exceeds an upper limit. 
As a result, we can expect the following scenario for the evolution of the black hole mass: 
When the black hole mass is smaller than the upper limit, mass accretion of the interstellar matter in the circum-nuclear region, causing the AGN activities, makes the black hole mass larger.
However, when the mass gets to the upper limit, the black hole loses the balancing force against the dynamical friction and starts spiraling downward to the gravity center. From simple parameter scaling, the upper mass limit of the black hole is found to be proportional to the bulge mass and this could explain the observed correlation of the black hole mass with the bulge mass.

\end{abstract}

\section{Introduction}
One of the most remarkable findings in astrophysics 
in the past a few decades should be 
the tight correlation of the central black hole mass to 
the velocity dispersion or mass of the bulge component in the host galaxy 
(e.g. Ferrarese \& Merritt 2000; Gebhardt et al. 2000; Merritt \& Ferrarese 2001; Marconi \& Hunt 2003).
This discovery has motivated various theoretical and observational studies, 
mainly focusing on the co-evolution of the black hole with the ambient galactic bulge 
through several feedback processes (see Fabian 2012; Kormendy \& Ho 2013; 
Heckman \& Best 2014, for the review, and references therein). 

These studies are based on a general consensus that 
activities of active galactic nuclei (AGN) are results of mass accretion on to the central black hole and 
that the accreted matter is supplied from the outside of the nucleus, which is triggered by 
such activities as star bursts, galactic mergers and so on, on the host galaxy side.

However, there still remains an important and difficult issue, 
how the accreted matter throws away its angular momentum and finally gets to the 
black hole at the center:
The processes for matter to inflow from a region with a $\sim$kpc distance 
to a region with a 10 $\sim$ 100 pc distance have been studied fairly well both observationally 
and theoretically, but the final path from the 10 - 100 pc region to an accretion-disk  region with a sub-pc distance to the black hole is still far being understood (see e.g. Jogee 2006; Alexander \& Hickox 2012, for the 
review and references therein).

Recently, Inoue (2020) argues that several observational properties of AGNs could well be  explained by considering a mass accretion caused by the passage of a black hole through the interstellar space in a circum-nuclear region.
If a black hole with the mass, $M_{\rm bh}$, passes with relative velocity, $v$, through the interstellar space with number density, $n_{\rm is}$, the mass accretion rate to the black hole, $\dot{M}$, is approximately given as 
\begin{equation}
\dot{M} \simeq \frac{4\pi (GM_{\rm bh})^{2} n_{\rm is} m_{\rm p}}{v^{3}},
\label{Mdot_iss}
\end{equation}
where $G$ and $m_{\rm p}$ are the gravitational constant and the proton mass, respectively (e.g. Davidson \& Ostriker 1973).
Then, the X-ray luminosity, $L_{\rm X}$ is estimated as
\begin{eqnarray}
L_{\rm X} &\simeq& \eta \dot{M} c^{2} \nonumber \\
&\simeq& 1 \times 10^{45} \left(\frac{\eta}{0.01}\right) \; \left(\frac{v}{10^{6.5}\rm{cm}}\right)^{-3} \; \left(\frac{n_{\rm is}}{10^{2}\; \rm{cm}^{-3}}\right)  \left(\frac{M_{\rm bh}}{10^{7}M_{\odot}}\right)^{2} \; \textrm{erg s}^{-1},
\label{eqn:L_X}
\end{eqnarray}
where $\eta$ is the energy conversion efficiency of the accretion onto the black hole.
If we consider a case in which a black hole with the mass of 10$^{7} M_{\odot}$ moves in such a circum-nuclear region as the nuclear stellar disk of our Galaxy (Launhardt, Zylka \& Mezger, 2002), we can adopt $n_{\rm is} \simeq 10^{2}$ cm$^{-3}$ and $v \simeq 10^{6.5}$ cm s$^{-1}$ as the typical values.  We further assume
 $\eta \simeq 0.01$ taking account of the low radiative efficiency of the accretion disks in AGNs (see e.g. Inoue 2020).   $L_{\rm X}$ estimated with those parameters roughly agree to the observed values.
Then, Inoue (2020) further points out a possibility that the central black hole might wander around the galactic center and sometimes get close to the circum-nuclear,  10 - 100 pc region in which dense interstellar matter could exist, referring to the recent observational report by Combes et al. (2019) that the AGN positions were frequently off-centered by several tens pc in the circumnuclear structures.

In this paper, we study a mechanism for the central black hole to move around the nuclear center and investigate how the black hole mass evolves there.
We propose a mechanism for the black hole to move and estimate how far it can go out from the center in section 2.
As a result, it is shown that the central black hole could be possible to come to the 10 - 100 pc region.
Finally, we discuss if the observed correlation between the black hole mass 
and the bulge mass could be explained in the scope of this scenario, in section 3.

\section{Wandering of the central black hole}
Wandering of the central massive object in a large gravitationally bound system 
has already been studied for a case of a cD galaxy in a cluster of galaxies by Inoue (2014), where a situation is considered in which member galaxies come closest to 
the cluster center in turn and cause the central object to be pulled 
to random directions one after another.
The encounter between the central object and the innermost galaxy induces a movement of the central object, trying to establish the equipartition of their kinetic energies between them.
If the central object starts moving, however, 
the ambient dark matter particles brake the moving through the dynamical friction.
Then, Inoue (2014) approximately calculated the energy gain rate of the central object 
from the innermost galaxy passing by the central object and the energy loss rate 
through the dynamical friction to the diffuse dark matter, 
and obtained the wandering velocity of the central object 
by balancing the energy gain rate and the energy loss rate.

Adopting a different situation to that of Inoue (2014), we now consider the case that a super-massive black hole stays at the center of a galactic bulge which consists of 
a large number of stars, loading most of the bulge mass, and interstellar gas including a certain number of molecular clouds.
The study by Inoue (2014) can be applied to the present case, 
by replacing the central cD galaxy,  
member galaxies and diffuse dark matter, in a cluster of galaxy, with a central black hole, molecular clouds and field stars, in a bulge, respectively.

\subsection{Environments of the super-massive black hole}\label{Environments}
Following the case of our Galaxy studied by Launhardt, Zylka \& Mezger (2002),
a nuclear bulge is supposed to surround a super-massive  black hole with a mass, $M_{\rm bh}$, and to consist of a nuclear stellar disk (NSD) and a nuclear stellar cluster (NSC).
Although the NSD should have a spheroidal shape, we approximate its structure to be spherically symmetric with a radius, $R_{\rm d}$ and the total mass, $M_{\rm d,\; t}$, for simplicity.
The density in the NSD, $\rho_{\rm d}$, is assumed to be constant, according to the phenomenological result on the density distribution of the inner part of the Galactic NSD  (Launhardt, Zylka \& Mezger, 2002).
Thus, we have the relation as
\begin{equation}
M_{\rm d,\; t} = \frac{4\pi}{3} \rho_{d} R_{\rm d}^{3}.
\label{eqn:M_d,t-R_d}
\end{equation}
The NSC is also assumed to have a spherical structure with radius, $R_{\rm c}$, and total mass, $M_{\rm c,\; t}$.
We further assume 
\begin{equation}
M_{\rm d,\; t} \simeq 10^{2}\; M_{\rm c,\; t},
\label{eqn:M_d-M_c}
\end{equation}
and
\begin{equation}
R_{\rm d} \simeq 10\; R_{\rm c},
\label{eqn:R_d-R_c}
\end{equation}
following the case of the Galaxy again.

Several percent of the NSD mass is the interstellar matter and its significant fraction could be in dense, clumpy regions called as molecular clouds.
The molecular clouds in the NSD are assumed to have a mass spectrum as
\begin{equation}
\frac{dN}{dM_{\rm mc}} = 
\frac{N_{\rm u}}{M_{\rm mc,\; u}} \left( \frac{M_{\rm mc}}{M_{\rm mc,\; u}}\right)^{-\gamma} \mbox{ for } M_{\rm mc} \leq M_{\rm mc,\; u} 
\label{eqn:dNdM}
\end{equation}
where $M_{\rm mc}$ is the mass of a molecular cloud, $N(M_{\rm mc})$ is the number of the molecular clouds with the mass $\leq M_{\rm mc}$ in the NSD, $M_{\rm mc,\; u}$ is the upper mass limit and $N_{\rm u}$ is the normalization number.
From this equation, we get the total mass of the molecular clouds, $M_{\rm mc, t}$, in the NSD as
\begin{equation}
M_{\rm mc,t} = \frac{1}{2-\gamma}N_{\rm u} M_{\rm mc,\; u},
\label{eqn:M_mc,t}
\end{equation}
when $\gamma < 2$.
The mass spectral index, $\gamma$, of the several nearby galaxies including the galactic center region of our Galaxy are observed to be $\sim$ 1.5 $\sim$ 1.6 (Miyazaki \& Tsuboi, 2000; Rosolowsky 2005; Fukui \& Kawamura, 2010).

In the following subsections, we evaluate the wandering velocity of a black hole in the NSD by equating the energy gain rate of the black hole through gravitational attractions from the nearby molecular clouds in the NSD with the energy loss rate through dynamical frictions from the ambient stars again in the NSD.
This situation should raise an energy flow from the molecular clouds via the black hole to the field stars in the NSD, and thus the related properties of the NSD are regarded to be time variable.
The fractional variation-amplitudes are, however, expected to be negligibly small, since the total stellar mass and the total molecular mass are both considerably larger than that of the black hole.
Thus, we hereafter assume that properties of the NSD and NSC are all constant 
over a time in which the black hole mass evolves in the NSD.
We do not consider such activities inside the nuclear bulge as star burst activities nor effects on the nuclear bulge of large scale activities in the host galaxy.

\subsection{Gravitational attractions of the molecular clouds on the black hole}
\label{GravitationalAttractions}
Attractions of the molecular clouds to the black hole are approximately estimated in appendix \ref{dEdt_mc-bh}.
From the result in equation (\ref{eqn:dE/dt_final}), 
the energy flow rate to the black hole with wandering velocity, $u$, 
from the nearest molecular clouds with a mass spectrum in equation (\ref{eqn:dNdM}) and velocity, $v$, can be expressed as
\begin{eqnarray} 
\left. \frac{dE}{dt} \right |_{\rm mc-bh} &\simeq& \alpha M_{\rm bh} \frac{G^{2}}{ R_{\rm d}^{3} v} \int_{M_{\rm mc,\; e}}^{M_{\rm mc,\; u}} M_{\rm mc}^{2} \frac{dN}{dM_{\rm mc}} \; dM_{\rm mc}, \nonumber \\
&\simeq& \alpha \frac{2-\gamma}{3-\gamma} M_{\rm bh} \frac{G^{2} M_{\rm mc, \; t}\; M_{\rm mc,\; u}}{ R_{\rm d}^{3} v}, 
\label{eqn:dEdt_mc-bh}
\end{eqnarray} 
Here, $r_{1}^{(i)}$ in equation (\ref{eqn:dE/dt_final}) has been replaced with the average radius, $r_{1}$, of the spherical volume in which only one molecular cloud in a minute mass range between $M_{\rm mc}$ and $M_{\rm mc}+\Delta M$ exists on average and 
it has been approximated by an equation as
\begin{equation}
r_{1}^{3} \simeq \frac{R_{\rm d}^{3}}{(dN/dM_{\rm mc})\Delta M}.
\label{eqn:r_1}
\end{equation}
$\alpha$ represents the term of $\beta_{2}/(2\beta_{1}^{2})$ in equation (\ref{eqn:dE/dt_final}).
As mentioned in appendix \ref{dEdt_mc-bh}, $\beta_{1}$ is considered to be less than unity and hence it makes $\alpha$ larger than unity.
$\alpha$ should also include the deviation factor of the $r_{1}$ value in the spherically symmetric case from the real spheroidal case, and it could be larger than unity too.
Taking account of these factors, $\alpha$ is likely to be significantly larger than unity.
The lower limit mass, $M_{\rm mc,\; e}$, in the integration is the mass of the molecular cloud in the equipartition of the kinetic energy with the black hole, and is defined with  the following equation as,
\begin{equation}
M_{\rm mc,\; e} \frac{v^{2}}{2} = M_{\rm bh}\frac{u^{2}}{2}.
\label{eqn:M_mc,e}
\end{equation}
Molecular clouds with the mass larger than $M_{\rm mc,\; e}$ only contribute to the energy transfer from the molecular clouds to the black hole.
Considering $v^{2} \gg u^{2}$ in the present case, we practically set $M_{\rm mc,\; e} \sim$ 0.

\subsection{Dynamical frictions of the ambient stars on the black hole}
If the black hole moves with velocity, $u$, against the field stars in the NSD, 
it should get the dynamical friction force, $F_{\rm df}$, calculated as
\begin{equation}
F_{\rm df} = \frac{4\pi (GM_{\rm bh})^{2} \rho_{\rm d} \ln \Lambda}{u^{2}} A(X),
\label{eqn:F_df}
\end{equation}
where
\begin{equation}
A(X) = \rm{erf}(X) -\frac{2X}{\sqrt{\pi}}\rm{e}^{-X^{2}}
\label{eqn:A(X)}
\end{equation}
(Binney \& Tremaine, 2008). 
Here, $X$ is defined as
\begin{equation}
X = u/(\sqrt{2}\sigma), 
\label{eqn:X-Def}
\end{equation}
and $\sigma$ is the velocity dispersion of the stars.
According to Binney \& Tremaine (2008), the factor, $\Lambda$, in the Coulomb logarithm is roughly given to be $(M_{\rm d,\; t}/M_{\rm bh})(r_{\rm bh}/R_{\rm d})$, 
where $r_{\rm bh}$ is the orbital radius of the black hole around the NSD center.
When $M_{\rm bh} \simeq 10^{-2}\; M_{\rm d,\; t}$ and $r_{\rm bh}$ is of the order of $0.1\; R_{\rm d}$, $\ln \Lambda$ is around 3.
From this equation, the energy transfer rate from the black hole to the ambient stars through the dynamical friction, $dE/dt|_{\rm bh-st}$ is
\begin{eqnarray}
\left. \frac{dE}{dt} \right|_{\rm bh-st} &\simeq& F_{\rm df} \; u \nonumber \\
&\simeq& \frac{4\pi (GM_{\rm bh})^{2} \rho_{\rm d} \ln \Lambda}{\sqrt{2}\sigma} \frac{A(X)}{X}.
\label{eqn:dEdt_bh-st}
\end{eqnarray}

\subsection{Equilibrium position of the black hole}
The equation for the balance between the energy gain rate and the energy loss rate 
of the central black hole is given as
\begin{equation}
\left. \frac{dE}{dt} \right|_{\rm mc-bh} = \left. \frac{dE}{dt} \right|_{\rm bh-st},
\label{eqn:EnergyBalance}
\end{equation}
and this yields the following relation with helps of equations (\ref{eqn:dEdt_mc-bh}), (\ref{eqn:dEdt_bh-st}) and (\ref{eqn:M_d,t-R_d}) as
\begin{equation}
\frac{\sqrt{2}\alpha}{3\ln \Lambda} \frac{\sigma}{v} \frac{2-\gamma}{3-\gamma} \frac{M_{\rm mc,\; u}}{M_{\rm mc,\; t}} \left(\frac{M_{\rm mc,\; t}}{ M_{\rm d,\; t}}\right)^{2} \simeq \frac{M_{\rm bh}}{M_{\rm d,\; t}} \; \frac{A(X)}{X}.
\label{eqn:BalanceEq}
\end{equation}

The value of $X$ is calculated in appendix 2 as functions of the normalized distance of the black hole to the nucleus center, $x$, defined in equation (\ref{eqn:x}) and the relative mass of the black hole to the total nuclear bulge mass, $m_{\rm bh}$, defined as
\begin{equation}
m_{\rm bh} = \frac{M_{\rm bh}}{M_{\rm n,\; t}} \simeq \frac{M_{\rm bh}}{M_{\rm d,\; t}} ,
\label{eqn:m_bh}
\end{equation}
where $M_{\rm n,\; t}$ is the total nuclear bulge mass given by $M_{\rm n,\; t} = M_{\rm c,\; t} + M_{\rm d\; t}$.
We can now see the right side of equation (\ref{eqn:BalanceEq}) as functions of $x$ and $m_{\rm bh}$ and rewrite it as
\begin{equation}
B = C(x, m_{\rm bh}),
\label{eqn:B-C}
\end{equation}
where
\begin{equation}
B = \left(\frac{\sqrt{2}\alpha}{3\ln \Lambda} \frac{\sigma}{v}\right) \left(\frac{2-\gamma}{3-\gamma} \frac{M_{\rm mc,\; u}}{M_{\rm mc,\; t}}\right) \left(\frac{M_{\rm mc,\; t}}{ M_{\rm d,\; t}}\right)^{2}, 
\label{eqn:B_def}
\end{equation}
\begin{equation}
C(x, m_{\rm bh}) = m_{\rm bh} \frac{A(X)}{X},
\label{eqn:C}
\end{equation}
and $X = X(x, m_{\rm bh})$.
If $m_{\rm bh}$ is given, we can determine the equilibrium position of the black hole from this equation.

Figure 1 shows the C-values as a function of $x$ for six given $m_{\rm bh}$ values from $10^{-3.5}$ to $10^{-1}$.
The line of $B=1 \times 10^{-4}$ is drawn in the figure to show a case in which the $C$ curve for $m_{\rm bh} = 10^{-2}$ is tangential to the $B$ line at the position with the open circle.
We see from this figure that when $m_{\rm bh}$ is $10^{-3.5}$, $10^{-3}$ and  $10^{-2.5}$, the equilibrium position of the black hole gets $\sim 0.8$, $\sim 0.5$ and $\sim 0.3$ respectively.
Since the typical size of the NSD, $R_{\rm d}$, could be $\sim 100$ pc, we can say that the black hole could wander in the NSD region at a distance of several 10 pc from the nucleus center in these cases.

\begin{figure}
 \begin{center}
  \includegraphics[width=10cm]{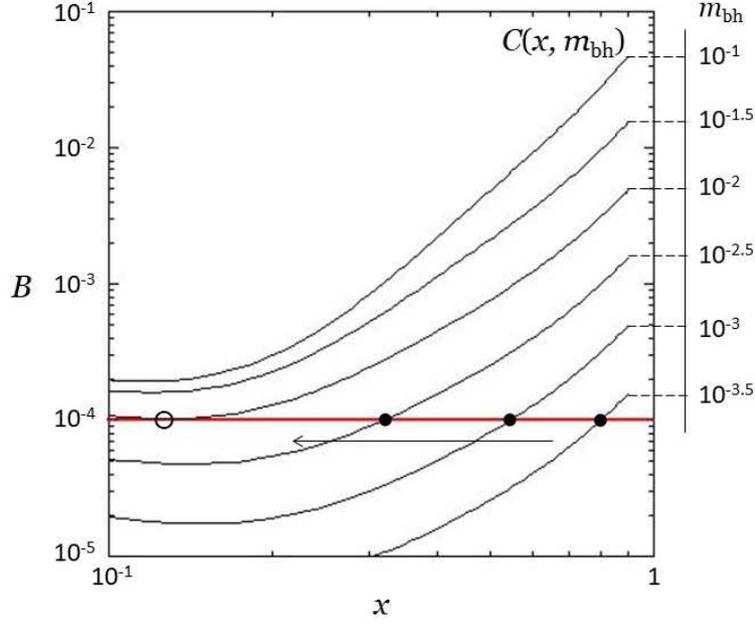} 
 \end{center}
\caption{Examples of a constant $B$ line and $C$ curves, which determine the equilibrium position of the black hole between the gravitational attractions from the nearby molecular clouds and the dynamical friction from the ambient field stars in the nuclear stellar disk. The straight line corresponds to the case of $B=1 \times 10^{-4}$, and the curves represent the $C$-values as functions of $x$ in six cases of $m_{\rm bh}$.  The cross point between the $B$-line and the $C$-curve exhibits the equilibrium position of the black hole for given $B$ and $m_{\rm bh}$ values.  The arrow shows the movement of the equilibrium point associated with the mass increase of the black hole.  The open circle at the position where the $B$ line and the $C$-curve touches with each other indicates the upper limit mass of the black hole for the given $B$ value. }\label{Figure1}
\end{figure}

\section{Evolution of the black hole mass}
If the above scenario really takes place, we can expect that a black hole with an appropriate mass wanders in the NSD region.
Then, since the NSD region contains a significant amount of interstellar matter, matter accretion on to the black hole is expected to happen during the passage of the black hole through the interstellar space and this makes the black hole mass larger.
As seen from figure 1, the position of the black hole moves inward as the black hole mass increases, and eventually gets to the position where the curve C touches to the line B.
Since mass accretion continues even at this position, the black hole mass further increases and the black hole goes into a situation in which the B and C lines do not cross each other any more.
Now that the dynamical friction is always dominant to the attractions from the molecular clouds, it continuously forces the black hole in-fall.
Since the rotational velocity of the black hole increases as the radius decreases as seen in figure 2, the dynamical friction gets stronger and stronger, associated with the in-fall in this situation.
Furthermore, the stellar density of the NSC is much higher than that of the NSD and thus the dynamical friction could be much strengthened after the black hole enters the NSC region.
As a result, the black hole is expected to spiral downward to the nucleus center in a fairly short time after it reaches to the tangential point between the B and C lines.

The above arguments indicate that a black hole with a relatively small mass increases its mass through mass accretion caused by wandering in the NSD, but that no significant mass increase occurs after it reaches the position where the B and C lines touches with each other.
Thus, the presence of an upper mass limit of the black hole is predicted from the present scenario.

It should be noted here that the key parameters introduced in the above arguments are several ratios of pairs of masses or radii.
They are $M_{\rm d,\; t}$ - $M_{\rm c,\; t}$ and $R_{\rm d}$ - $R_{\rm c}$ for the nuclear bulge, $M_{\rm mc,\; u}$ - $M_{\rm mc,\; t}$ for the mass spectrum of the molecular clouds, $M_{\rm mc,\; t}$ - $M_{\rm d,\; t}$ for the mass fraction of the molecular clouds in the NSD, and $M_{\rm bh}$ - $M_{\rm n,\; t}$ for the mass ratio of the black hole to the nuclear bulge.
Hence, if these ratios are universal to every galaxies, 
the observed proportionality between the black hole mass and the bulge mass which is probably proportional to the nuclear bulge mass can be explained with this scenario.

In order for the tangential point between the B and C lines to appear, the shape of the C curve should be convex downward and it needs the presence of the NSC having the appropriate mass and radius relative to those of the NSD.
As seen in figure 2, the profile of $u_{\rm k}^{2}$ has the minimum where $x$ is close to 0.1 and this makes the C-curve convex downward.
The gravity by the NSD matter within $x$ is dominant to that of the NSC in the region far from the NSC and $u_{\rm k}^{2}$ is proportional to $x^{2}$. 
However, the gravity from the NSC becomes dominant in the region near the NSC and $u_{\rm k}^{2}$ tends to be proportional to $x^{-1}$.
This situation causes the appearance of the minimum in $u_{\rm k}^{2}$.
Although the ratios in equations (\ref{eqn:M_d-M_c}) and (\ref{eqn:R_d-R_c}) are based on the observations of the Galaxy (Launhardt, Zylka \& Mezger, 2002) and those informations from other galaxies are poor yet, Ferrarese et al. (2006) reports that the centers of most galaxies are occupied by the NSCs, and suggests the tight correlation of the NSC masses with the masses of the host galaxies.

The ratio between $M_{\rm mc,\; u}$ and $M_{\rm mc,\; t}$ appears, together with the index, $\gamma$, of the mass spectrum of the molecular clouds, in equation (\ref{eqn:B_def}) to determine the B value.
$M_{\rm mc,\; u}$ and $N_{\rm u}$ are estimated to be roughly $10^{6.5} M_{\odot}$ and $10^{0.5}$ from the mass spectrum of the molecular clouds in the Galactic center region obtained by Miyazaki \& Tsuboi (2002).
From these values and equation (\ref{eqn:M_mc,t}) adopting $\gamma \sim 1.5$,  we get $M_{\rm mc,\; t} \sim 10^{7.3} M_{\odot}$ and an estimation for the second parenthesis in equation (\ref{eqn:B_def}) of
\begin{equation}
\frac{2-\gamma}{3-\gamma} \frac{M_{\rm mc,\; u}}{M_{\rm mc,\; t}} \sim 10^{-1.3}.
\label{eqn:2ndP_B}
\end{equation}
The ratio of $M_{\rm mc,\; t}$ to $M_{\rm d,\; t}$ is also included in the estimation of $B$.
Since $M_{\rm d,\; t}$ is reported to be $\sim 10^{9} M_{\odot}$ in the case of the Galaxy (Launhardt, Zylka \& Mezger, 2002), we have
\begin{equation}
\frac{M_{\rm mc,\; t}}{M_{\rm d,\; t}} \sim 10^{-1.7}.
\label{eqn:M_mc,t-M_d,t}
\end{equation}
The molecular gas content per unit stellar mass is shown to be fairly constant for all late-type galaxies of type Sa to Im and is about 10\% (Boselli et al. 2014).
Here, it is set to be about 2\%, considering that our concern is on the nuclear bulge and the molecular clouds.
We further need to set the value of $\alpha$ in the right side of equation (\ref{eqn:B_def}) to identify the value of $B$; 
it could be as large as ten or so as discussed in subsection \ref{GravitationalAttractions}.
Hence, we assume $\alpha \sim 10$ and then have roughly
\begin{equation}
\frac{\sqrt{2}\alpha}{3\ln \Lambda} \frac{\sigma}{v} \sim 10^{0.2},
\label{eqn:1stP-B}
\end{equation}
considering $\sigma \sim v$ and $\ln \Lambda \sim 3$.
Inserting the parameter values evaluated above, $B$ is approximately estimated as
\begin{equation}
B \sim 10^{-4.5}.
\label{eqn:B-FinalValue}
\end{equation}

As seen from figure 1, the upper limit of $m_{\rm bh}$ is slightly lower than  $\sim10^{-2.5}$ for $B \simeq 10^{-4.5}$.
Thus, the present scenario predicts that the ratio of the upper mass limit of the black hole to the bulge mass could be a little less than $10^{-3.5}$,
if the mass of the nuclear bulge is about 10\% of that of the galactic bulge.
This is smaller than the observed mass ratio $\sim 10^{-3}$ (Merritt \& Ferrarese 2001; Maeconi \& Hunt 2003) by about a factor of 5 or so, but the difference could be within the ambiguities in the order estimations of the present study.
 

\section{Summary}
The starting point of the study in this paper is
the recent argument by Inoue (2020) that
a passage of the super-massive black hole through interstellar matter in a circum-nuclear region could be an origin of AGN activities.
Even if this hypothesis is correct, however, a serious question arises:  
How can the interstellar matte in the circum-nuclear region get close to the black hole at the nucleus center?
No direct link has yet been found observationally or theoretically between the nuclear region with the 10 $\sim$ 100 pc distance and the accretion-disk region with a sub-pc distance.

Reversing the way of thinking, here, we have investigated 
a possibility that the black hole wanders around the nucleus center and comes close to the outer nuclear region, rather than the matter in the circum-nuclear region going down to the center. 
A situation has been considered in which a certain number of molecular clouds move around in a nuclear bulge and the central black hole is pulled by nearby molecular clouds.
We have estimated the balance between the energy transfer rate from the molecular clouds to the black hole through the gravitational acceleration from the nearby molecular clouds and that from the black hole to the ambient stars through the dynamical friction from the ambient stars in the nuclear bulge.
Then, it is found that the black hole can have enough kinetic energy to wander in a region with a radius as large as several 10 pc, when the black hole mass is relatively small.  At the same time, it is predicted that a black hole with the relatively small mass increases its mass through mass accretion but loses its stable orbit when its mass exceeds the critical mass.

The studies carried out in this paper include several bold assumptions, approximations and simplifications that should be checked more precisely.
In spite of this, the proposed scenario of a wandering  
black hole in the circum-nuclear region at a distance 
of several 10 pc is very attractive in the following two ways.

One is that it is possible to give an answer to the difficult issue, which has long been studied in various ways but is yet unresolved, of
how the matter in the 10 - 100 pc region can reach 
the vicinity of the black hole and cause AGN activities.
Studies have currently been based on the thought that the matter in the 10 - 100 pc region should be flowing further inward to the sub-pc region of the accretion disk and have been faced with the difficult problems of the barriers of the angular momentum or the star formation.
The present scenario does not need to break those hard barriers.

The other is that it is possible to provide the simple interpretation of 
another difficult issue; the correlation of the black hole mass with the ambient bulge mass.
With this interpretation, the correlation between the two masses can be understood as 
a mere result of the whole accretion histories of the black hole, irrespectively of what has been happening in interactions between the black hole and the host galaxy, such as co-evolution or feedback.

The present scenario, which provides possible and fairly simple interpretations of these difficult issues, should be worthy of further study both observationally and theoretically.

\begin{ack}
The previous version of this manuscript was submitted to PASJ once but withdrawn since the author could not respond to the critical comment from the referee in a timely manner.
The present version is the result of reconsidering the point from the referee deeply.  The author is very grateful to the previous referee for the useful comment.
The referee to the present version is also appreciated for the helpful comments.
\end{ack}

\appendix 

\section{Energy transfer rate from the molecular clouds to the black hole}\label{dEdt_mc-bh}
We consider a situation that molecular clouds in a nuclear bulge consist of $I$ species having different masses, $M_{\rm mc}^{(i)}$, and number in the nuclear bulge,  $N^{(i)}$, ($i = 1,..., I$), from one another.

Defining $r_{1}^{(i)}$ as the radius of a sphere 
in which only one $i$-th the molecular cloud 
exists on average, we further introduce an assumption that 
all the gravitational forces from the $i$-th molecular clouds other than that from the nearest one are completely canceled by one another and that 
the black hole receives only a force from the nearest one. 
The force per mass, $f^{(i)}$, can be approximated as 
\begin{equation}
f^{(i)} \simeq \frac{GM_{\rm mc}^{(i)}}{\beta_{1} (r_{1}^{(i)})^{2}},
\label{eqn:f^(i)}
\end{equation}
where $\beta_{1}$ is the averaging factor of the distance between the nearest molecular cloud and the black hole during the passage of one particular nearest cloud and should be less than unity.
One among the molecular clouds in the $i$-th specie enters the spherical region with the radius $r_{1}^{(i)}$ one after another, and the average passage time, $\Delta t^{(i)}$, can be expressed as
\begin{equation}
\Delta t^{(i)} \simeq \beta_{2} \frac{r_{1}^{(i)}}{v},
\label{eqn:Delta-t^i}
\end{equation}
where $\beta_{2}$ is another averaging factor over various passing orbits, of the order of 1.

Let us introduce a three-dimensional diagonal-coordinate ($x$, $y$, $z$), and express the three axis components of the specific gravitational force which the black hole gets from the $i$-th  molecular clouds as $f_{k}^{(i)}$ ($k$ = $x$, $y$, $z$).
Then, the velocity of the black hole, $u$, at time, $t$, can be calculated as 
\begin{eqnarray}
u_{k} &=& \int_{0}^{t} \sum_{i = 1}^{I} f_{k}^{(i)} dt' \nonumber \\
&\simeq& \sum_{i=1}^{I} \sum_{j=1}^{J^{(i)}} (f_{k}^{(i)} \Delta t^{(i)})_{j}, 
\label{eqn:u_ell}
\end{eqnarray}
for $k$ = $x$, $y$, $z$.
Here, $j$ represents the sequence number of the successive replacements of the closest molecular cloud to the black hole, and its total number in $t$ is $J^{(i)} = \mbox{int}[t/\Delta t^{(i)}]$.
Since the direction of the velocity increment during each $\Delta t^{(i)}$ should be random, the average of $f_{k}^{(i)} \Delta t^{(i)}$ over a large number of $J^{(i)}$ should be zero, and thus the average of $u_{k}$ over a sufficiently long time of $t$ should become zero for all $x$, $y$ and $z$ in $k$.
The average of $u_{k}^{2}$, $\overline{ u_{k}^{2} }$, is, on the other hand, calculated as 
\begin{eqnarray}
\overline{ u_{k}^{2} } &\simeq& \overline{ \left(\sum_{i=1}^{I} \sum_{j=1}^{J^{(i)}} (f_{k}^{(i)} \Delta t^{(i)})_{j}\right)^{2}  } \nonumber \\
&\simeq& \sum_{i=1}^{I} \overline{ (f_{k}^{(i)} \Delta t^{(i)})^{2} },
\label{eqn:u_k^2}
\end{eqnarray}
since all the cross terms should become zero on average.
From equation (\ref{eqn:u_k^2}) and with the help of equations (\ref{eqn:f^(i)}) and (\ref{eqn:Delta-t^i}), we get the energy transfer rate from all the molecular clouds to the black hole, $dE/dt|_{\rm mc-bh}$, as
\begin{eqnarray}
\left. \frac{dE}{dt} \right|_{\rm mc-bh} &\simeq& \frac{M_{\rm bh}}{2} \left( \frac{d\overline{u_{\rm x}^{2}}}{dt} +  \frac{d\overline{u_{\rm y}^{2}}}{dt} + \frac{d\overline{u_{\rm z}^{2}}}{dt} \right) \nonumber \\
&\simeq&  \frac{M_{\rm bh}}{2} \sum_{i=1}^{I} \frac{\overline{ (f^{(i)} \Delta t^{(i)})^{2} }}{\Delta t^{(i)}} \nonumber \\
&\simeq& \frac{M_{\rm bh}}{2} \sum_{i=1}^{I} \frac{\beta_{2} (G M_{\rm mc}^{(i)})^{2}}{\beta_{1}^{2} (r_{1}^{(i)})^{3} v}.
\label{eqn:dE/dt_final}
\end{eqnarray}

\section{Kinematics of the black hole and the ambient stars in the NSD}\label{Kinematics}

\subsection{Black hole velocity}
We assume that the black hole circularly rotates around the gravity center of the NSD and black hole system.
Under the environmental conditions introduced in subsection \ref{Environments}, 
the mass $M_{\rm n}$ of the NSD + NSC matter within a distance, $r$ ($>R_{\rm c}$), from the NSD center is given as
\begin{eqnarray}
M_{\rm n} &=& \frac{4\pi \rho_{\rm d} }{3} \; (r^{3}-R_{\rm c}^{3}) + M_{\rm c,\; t} \nonumber \\
&=& M_{\rm d,\; t} \; (x^{3}-x_{\rm c}^{3}) + M_{\rm c,\; t},
\label{eqn:M_n-x}
\end{eqnarray}
where 
\begin{equation}
x=\frac{r}{R_{\rm d}},
\label{eqn:x}
\end{equation}
and 
\begin{equation}
x_{\rm c} = \frac{R_{\rm c}}{R_{\rm d}}.
\label{eqn:x_c}
\end{equation}
Then, the Keplerian circular velocity of the black hole, $u_{\rm k}$, is calculated as a function of $x$ by the following equation as
\begin{eqnarray}
u_{\rm k}^{2} &\simeq& D\; r\; \frac{GM_{\rm n}}{r^{2}} \nonumber \\
&\simeq& D\; 
\frac{GM_{\rm d,\; t}}{R_{\rm d}} \left[ x^{2}+\frac{(M_{\rm c,\; t}/M_{\rm d,\; t})-x_{\rm c}^{3}}{x} \right],
\label{eqn:u_k-x}
\end{eqnarray}
where $D$ is defined as
\begin{equation}
D = \frac{M_{\rm n}}{M_{\rm n}+M_{\rm bh}}.
\label{eqn:D}
\end{equation}
Since $M_{\rm n}$ is a function of $x$, $D$ is also a function of $x$.
This factor, $D$, comes from a simple expectation that the black hole tends to rotate around the gravity center between the NSD within $r$ and the black hole, although more detailed considerations on the situation of the whole NSD + black hole system is necessary in practice.

\subsection{Velocity dispersion of the stellar motions}
Applying the thermo-dynamics to the stellar system in the NSD, we introduce the pseudo-pressure of the stellar motions, $P$, as
\begin{equation}
P = \rho_{\rm d} w^{2},
\label{eqn:P}
\end{equation}
where $w$ is the pseudo sound velocity of the stellar system.
The dynamical structure could be obtained by an equation as
\begin{equation}
\frac{dP}{dr} = -\rho_{\rm d} \frac{GM_{\rm n}}{r^{2}},
\label{eqn:dPdr}
\end{equation}
and this equation yields the following equation, on the assumption of the constant $\rho_{\rm d}$, as
\begin{equation}
\frac{dw^{2}}{dr} = -\frac{GM_{\rm n}}{r^{2}}.
\label{eqn:dw^2dr}
\end{equation}
The solution of this differential equation is, with the help of equation (\ref{eqn:M_n-x}), 
\begin{equation}
w^{2} = w_{\rm c}^{2} - \frac{GM_{\rm d,\; t}}{R_{\rm d}} \left[ \frac{x^{2}}{2}-\frac{x_{\rm c}^{2}}{2} + \left(  \frac{M_{\rm c,\; t}}{M_{\rm d,\; t}} - x_{\rm c}^{3} \right) \left( \frac{1}{x_{\rm c}} - \frac{1}{x} \right) \right],
\label{eqn:w^2-0}
\end{equation}
where $w_{\rm c}$ is $w$ at $r = R_{\rm c}$.
If we set the outer boundary condition as $w=0$ at $r=R_{\rm d}$, 
$w_{\rm c}$ is calculated as
\begin{equation}
w_{\rm c}^{2} = \frac{GM_{\rm d,\; t}}{R_{\rm d}}  \left[ \frac{1}{2}-\frac{x_{\rm c}^{2}}{2} + \left(  \frac{M_{\rm c,\; t}}{M_{\rm d,\; t}} - x_{\rm c}^{3} \right) \left( \frac{1}{x_{\rm c}} - 1 \right) \right].
\label{eqn:w_c^2}
\end{equation}
From the above two equations, we get
\begin{equation}
w^{2} = \frac{GM_{\rm d,\; t}}{R_{\rm d}}  \left[ \frac{1}{2}-\frac{x^{2}}{2} + \left(  \frac{M_{\rm c,\; t}}{M_{\rm d,\; t}} - x_{\rm c}^{3} \right) \left( \frac{1}{x} - 1 \right) \right].
\label{eqn:w^2-final}
\end{equation}

From the analogy to the thermo-dynamics, $w^{2}$ could relate to $\sigma^{2}$ as
\begin{equation}
w^{2} = \frac{\sigma^{2}}{3}.
\label{eqn:w^2-sigma^2}
\end{equation}

\subsection{X - x relation}
Now, we can calculate $X$ defined in equation (\ref{eqn:X-Def}) as a function of $x$ from
\begin{equation}
X^{2} = \frac{u_{\rm k}^{2}}{2\sigma^{2}} = \frac{u_{\rm k}^{2}}{6 w^{2}},
\label{eqn:X^2-u^2/w^2}
\end{equation}
with equations (\ref{eqn:u_k-x}) and (\ref{eqn:w^2-final}).

$u_{\rm k}^{2}/(GM_{\rm d,\; t}/R_{\rm d})$, $w^{2}/(GM_{\rm d,\; t}/R_{\rm d})$ and $X$ in a case of $M_{\rm c,\; t}/M_{\rm d,\; t} = 10^{-2}$ and $x_{\rm c} = 10^{-1}$ are plotted as functions of $x$ in figure 2.

\begin{figure}
 \begin{center}
  \includegraphics[width=8cm]{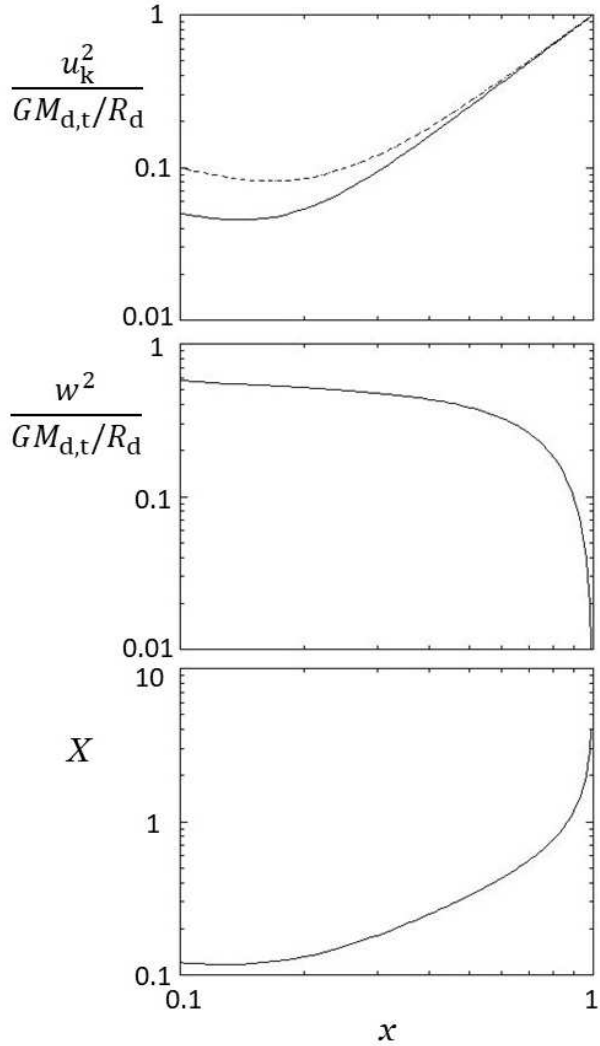} 
 \end{center}
\caption{(Top) The square of the expected rotational velocity, $u_{\rm k}$, of the black hole normalized by $GM_{\rm d,\; t}/R_{\rm d}$ as a function of $x$, (middle) the square of the pseudo-sound velocity of the stellar system, $w$, normalized by $GM_{\rm d,\; t}/R_{\rm d}$ as a function of $x$ and (bottom) $X$ as a function of $x$, in a case of $M_{\rm c,\; t}/M_{\rm d,\; t} = 10^{-2}$ and $x_{\rm c} = 10^{-1}$.  The solid and dashed curves in the top panel respectively correspond to the cases with and without the correction for the movement of the black hole around the gravity center between the black hole and the mass of the nuclear bulge within the black hole distance.   }\label{figure-2}
\end{figure}

\end{document}